\documentclass[a4paper,11pt]{article}
\usepackage{pos}
\usepackage{multirow}
\usepackage{mciteplus}
\title{Probing a light Higgs in $ H \to hh \to b\bar{b}\tau\tau$ in Type-I 2HDM at the LHC}

\author[a,b]{A. Arhrib}
\author[c,d]{S. Moretti}
\author*[c,e]{S. Semlali}
\author[e]{C.~H.~Shepherd-Themistocleous}
\author[f,g]{Y. Wang}
\author[h,i]{Q.S. Yan}

\affiliation[a]{Abdelmalek Essaadi University, Faculty of Sciences and Techniques, B.P. 2117 T\'etouan, Tanger, Morocco.}
\affiliation[b]{Department of Physics and Center for Theory and Computation, National Tsing Hua University, Hsinchu, Taiwan 300.}
\affiliation[c]{School of Physics and Astronomy, University of Southampton, Southampton, SO17 1BJ, United Kingdom.}
\affiliation[d]{Department of Physics and Astronomy, Uppsala University, Box 516, SE-751 20 Uppsala, Sweden.}
\affiliation[e]{Particle Physics Department, Rutherford Appleton Laboratory, Chilton, Didcot, Oxon OX11 0QX, United Kingdom.}
\affiliation[f]{College of Physics and Electronic Information, Inner Mongolia Normal University, Hohhot 010022, PR China.}
\affiliation[g]{Inner Mongolia Key Laboratory for Physics and Chemistry of Functional Materials, Inner Mongolia Normal University, Hohhot, 010022, China.}
\affiliation[h]{Center for Future High Energy Physics, Chinese Academy of Sciences, Beijing 100049, P.R. China.}
\affiliation[i]{School of Physics Sciences, University of Chinese Academy of Sciences, Beijing 100039, P.R. China.}

\emailAdd{aarhrib@uae.ac.ma}
\emailAdd{s.moretti@soton.ac.uk}
\emailAdd{stefano.moretti@physics.uu.se}
\emailAdd{souad.semlali@soton.ac.uk}
\emailAdd{claire.shepherd@stfc.ac.uk}
\emailAdd{wangyan@imnu.edu.cn}
\emailAdd{yanqishu@ucas.ac.cn}

\abstract{
	In this study, we explore the potential to probe the process $gg \to H_{\text{SM-like}} \to hh \to b\bar{b}\tau\tau$ at the Large Hadron Collider (LHC), within the framework of the Two Higgs Doublet Model (2HDM) Type-I. After performing a detailed Monte Carlo (MC) simulation, we focus on isolating the signal from the Standard Model (SM) backgrounds. Our analysis employs a dedicated trigger choice and optimised kinematic selection to improve the signal sensitivity. We demonstrate some sensitivity to this decay channel at Run 3, while the High-Luminosity LHC (HL-LHC) can provide discovery evidence. }

\FullConference{42nd International Conference on High Energy Physics (ICHEP2024)\\
18-24 July 2024\\
Prague, Czech Republic\\}

\begin{document}
\maketitle

\section{Introduction}

The recent measurements of the Higgs boson properties at the Large Hadron Collider (LHC)~\cite{ATLAS:2022vkf,CMS:2022dwd} align well with Standard Model (SM) predictions. However, given the current precision of the Higgs data, the possibility of non-SM Higgs decays remains open, with no conclusive evidence to rule them out. Both
ATLAS  \cite{ATLAS:2022vkf} and CMS \cite{CMS:2022dwd} experiments have set upper limits of 12\% and 16\% at 95\% C.L on the branching ratio of such non-SM decays. 

In light of this, the Two Higgs Doublet Model (2HDM), an extension of the SM that introduces an additional scalar sector, offers a promising framework to explore these potential deviations. One interesting process involves the exotic Higgs decay $gg \to H_{\text{SM-like}} \to hh \to b\bar{b}\tau\tau$, which has been actively investigated by the CMS experiment \cite{CMS:2024uru}, leading to constraints on its decay rate, while providing an alternative approach to probe Higgs self coupling. Our study focuses on the Type-I 2HDM to examine the feasibility of observing this decay channel, based on a signal-versus-background analysis using standard Monte Carlo (MC) simulation tools, at Run 3 and  High Luminosity LHC.

\section{2HDM Type-I}
The most general invariant scalar potential under ${SU(2)_L \times U(1)_Y}$ is given by:
\begin{eqnarray}
V_{\rm{Higgs}}(\Phi_1,\Phi_2) &=& \lambda_1(\Phi_1^\dagger\Phi_1)^2 +
\lambda_2(\Phi_2^\dagger\Phi_2)^2 +
\lambda_3(\Phi_1^\dagger\Phi_1)(\Phi_2^\dagger\Phi_2) +
\lambda_4(\Phi_1^\dagger\Phi_2)(\Phi_2^\dagger\Phi_1)  + \nonumber\\ && +
\frac12\left[\lambda_5(\Phi_1^\dagger\Phi_2)^2 +\rm{h.c.}\right] 
+m_{11}^2 \Phi_1^\dagger \Phi_1+ m_{22}^2\Phi_2^\dagger
\Phi_2 + \left[m_{12}^2
\Phi_1^\dagger \Phi_2 - \rm{h.c.}\right] \,.  \label{2hdmpot}
\end{eqnarray}
Due to the hermiticity of the scalar potential, the parameters $\lambda_{1,2,3,4}$, along with $m_{11}^2$ and $m_{22}^2$, are all real. $\lambda_5$ and $m_{12}^2$, on the other hand, can be complex and could potentially introduce CP violation in the Higgs sector. Herein, we assume CP conserving 2HDM.

Following Electroweak Symmetry Breaking (EWSB), the scalar sector consists of a pair of charged Higgs bosons ($H^\pm$), a CP-odd Higgs ($A$), and two CP-even Higgs bosons ($H$ and $h$) with $m_h < m_H$. One of the neutral CP-even states must correspond to the 125 GeV Higgs particle observed at the LHC. In this analysis, we assume $m_H = 125$ GeV, with $m_h < m_H/2$. The 2HDM Higgs sector can be described in the physical basis  by: 
\begin{eqnarray}
m_{H^{\pm}},~m_{A},~m_{H},~m_{h},~\sin \alpha,~\tan \beta=v_2/v_1~\text{and}~\ m_{12}^2,
\label{eq:param} 
\end{eqnarray}

The parameter space of the Two Higgs Doublet Model (2HDM) is thoroughly examined against various theoretical and experimental constraints. From a theoretical perspective, all Higgs potential parameters are required to satisfy conditions of unitarity, vacuum stability, and perturbativity. These theoretical constraints have been checked using the publicly available code \texttt{2HDMC-1.8.0}\cite{Eriksson:2009ws}. On the experimental side, we require compliance with EW precision tests [7]. To evaluate constraints from Higgs searches, signal strength measurements and flavour physics, we use \texttt{HiggsBounds-5.10.0} \cite{Bechtle:2020pkv}, \texttt{HiggsSignals-2.6.2} \cite{Bechtle:2020uwn} and \texttt{SuperIso}~\cite{Mahmoudi:2008tp}, respectively.

Focusing on the inverted hierarchy, we conducted a random scan over the following ranges,
\begin{center}
	$m_h$: $[10,~60]$\,GeV\,,~~$m_A$: $[62,~100]$\,GeV\,,~~$m_{H^{\pm}}$: $[100,~200]$\,GeV\,\\
	~~$s_{\beta - \alpha}$: $[-0.25,~-0.05]$,
	~~$m_{12}^2$: [0 -- $m_h^2\cos\beta \sin \beta$]\,,~~$\tan\beta$: [2 -- 25]\,. \\
\end{center} 

\begin{figure}[h!]	
	\hspace*{-0.3cm}
	\includegraphics[scale=0.4]{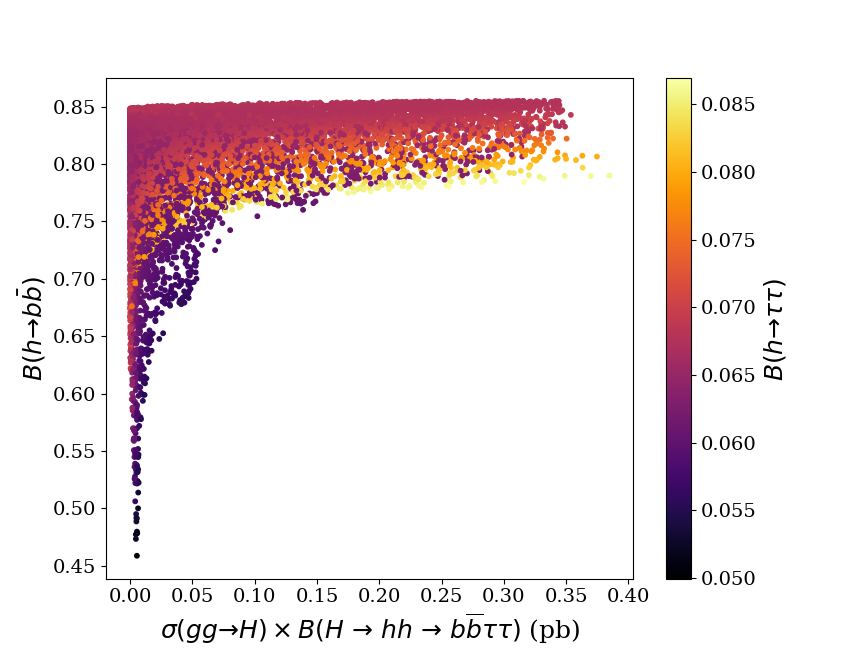}
	\includegraphics[scale=0.4]{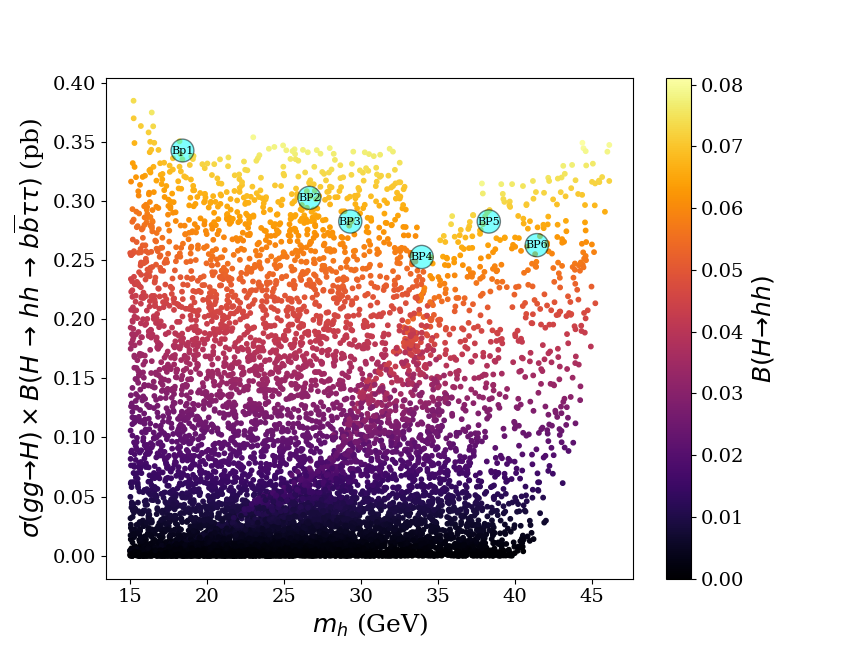}
	\caption{${\rm BR}(h \to b\overline{b})$ as a function of $\sigma(gg \to H \to hh \to b\overline{b}\tau\tau)$ vs. ${\rm BR}(h \to \tau \tau)$ (left panel).  $\sigma(gg \to H \to hh \to b\overline{b}\tau\tau)$ as a function of $m_h$ vs. ${\rm BR}(H \to hh)$ (right panel).}
	\label{fig2}
\end{figure}

Figure~\ref{fig2} shows the allowed points after scanning the 2HDM parameter space against the theoretical and experimental constraints discussed above. The plot highlights the cross-section of the process $pp \to H \to hh \to b\overline{b}\tau\tau$, where the SM-like Higgs boson $H$ is predominantly produced via gluon-gluon fusion. The cross-section for $gg \to H \to hh \to b\overline{b}\tau\tau$ reaches a maximum of 0.4 pb when the branching ratios ${\rm BR}(h \to b\overline{b})$, ${\rm BR}(h \to \tau\tau)$, and ${\rm BR}(H \to hh)$ are at their highest values. It's important to note that the light Higgs width is dominated by $h \to b\overline{b}$ decays, while ${\rm BR}(h \to \tau\tau)$ is approximately 7\%, and ${\rm BR}(H \to hh)$ is below 8\%. In this favourable region of the parameter space, several Benchmark Points (BPs) suitable for Monte Carlo simulations have been identified and marked in the Figure.

\section{Collider Phenomenology}
   To perform this MC analysis, we used \texttt{MadGraph-v.3.4.2}~\cite{Alwall:2014hca} to generate signal and background events, which are then passed to \texttt{PYTHIA8}~\cite{Sjostrand:2006za} for parton showering, fragmentation, hadronization, and heavy flavor decays. The detector response is simulated using \texttt{Delphes-3.5.0}~\cite{deFavereau:2013fsa} with a standard CMS card.
   We only focus on $b\bar b\tau_e\tau_\mu $ final state, whereas $\tau_e\tau_e$ and $\tau_\mu \tau_\mu$ states  are neglected to remove  the large Drell-Yan background. $\tau_e$ and $\tau_\mu$ represent $\tau \to e \bar{\nu}_e \nu_\tau$ and $\tau \to \mu \bar{\nu}_\mu \nu_\tau$, respectively. The main background processes arise from  $ Z(\to \tau_e \tau_\mu)b\overline{b}$  and $t\overline{t}$.

	To improve the efficiency of the event generation, the following kinematic cuts are applied:
\begin{equation*}
p_T(b/l) > 10/5~\text{GeV},~E_T^{\rm miss} > 5~\text{GeV}, ~|\eta(b,l)|<2.5,~\Delta R(ll,bl,bb) > 0.3,~H_T<70~\text{GeV}.
\end{equation*}	
	
After the detector simulation, we select events containing two b-jets and two oppositely charged leptons of different flavours ($e^\mp\mu^\pm$) in the final state. Since both leading and subleading leptons in the signal events are soft, a significant loss of signal is expected. To address this limitation, we proposed implementing a new cross electron-muon trigger, which would improve sensitivity to low transverse momentum ($p_T$) leptons and help overcome current trigger limitations in searches for such events~\cite{Arhrib:2023apw}.

\begin{figure}[h!]
	\includegraphics[scale=0.4]{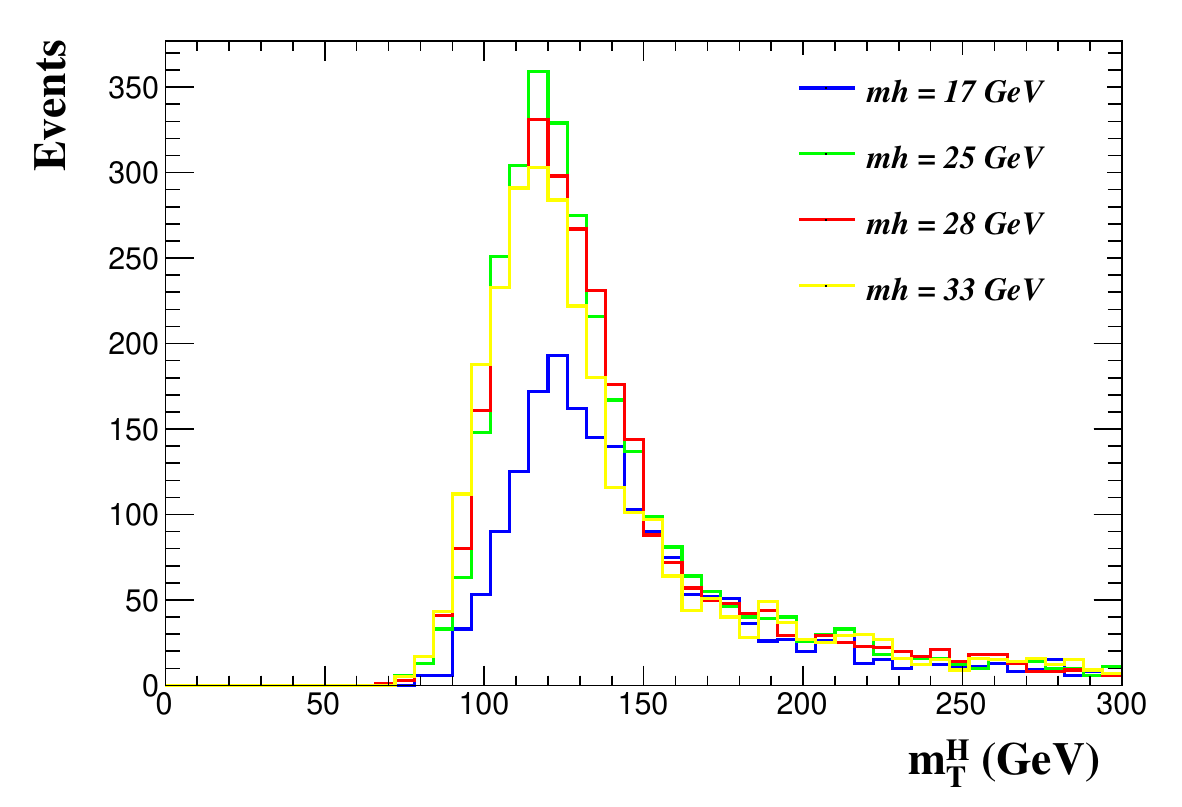}
	\includegraphics[scale=0.4]{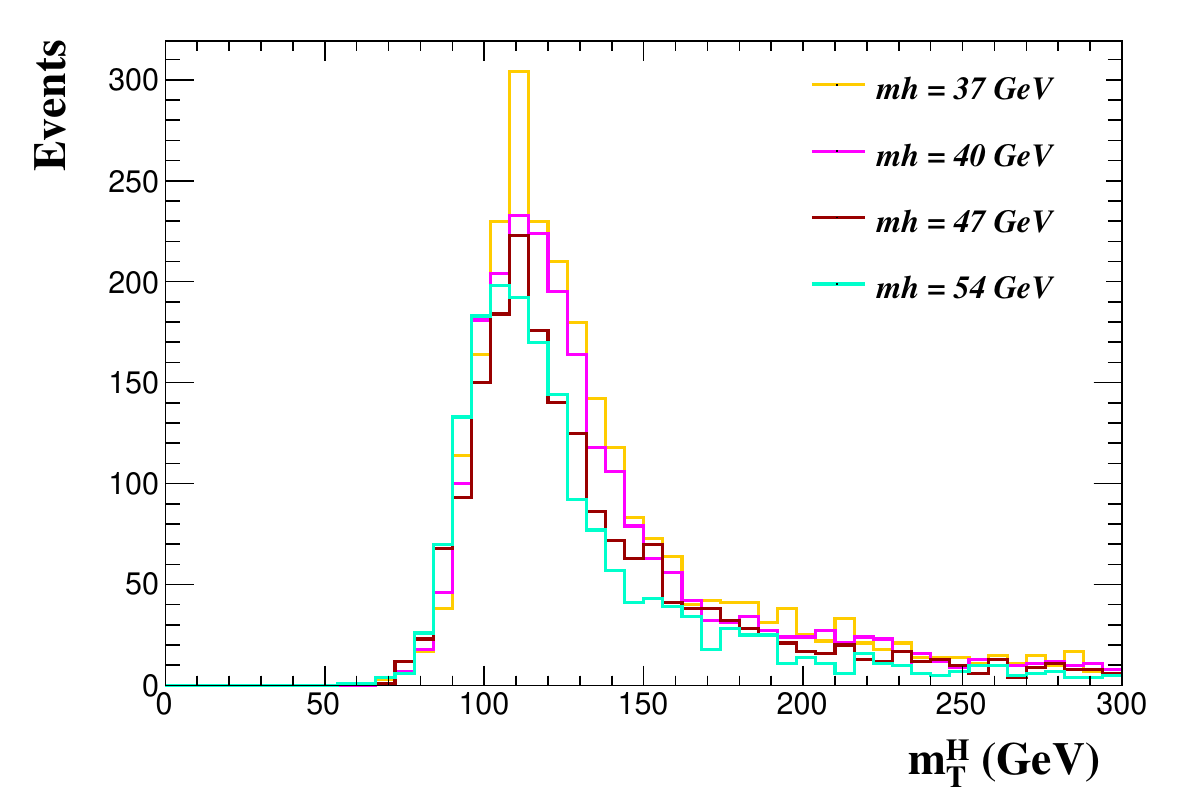}
	\caption{ The distributions of $m^{H}_T$ for different BPs are shown at detector level~\cite{Arhrib:2023apw}.}		
	\label{figa}	
\end{figure}

To improve the sensitivity of our analysis, we examine the transverse mass constructed from the $\tau$ lepton decay products and the $b$-jets ($m_T^H$), along with other relevant variables~\cite{Arhrib:2023apw}. These variables are expected to exhibit lower values for signal events, as they arise from a 125 GeV resonance, unlike the background events. Figure~\ref{figa} shows $m^{H}_T$ for different BPs,

Requiring exactly two $b$-tagged jets per event is a tight condition due to the low $p_T(b)$ and the limited $b$-tagging efficiency, especially in scenarios where the Higgs mass is below 60 GeV. To assess how the efficiencies, particularly $b$-tagging efficiency, are affected, we will evaluate the impact of three different cuts, which serve as pre-selection criteria:
\begin{equation}\label{btag-cuts}
p_T(b_1/b_2)>15/10~\text{GeV},~~~p_T(b_1/b_2)>20/15~\text{GeV},~~~p_T(b_1/b_2)>20/20~\text{GeV}.
\end{equation}	

The event rates for both signal and background are summarized in Tabs \ref{tab6} and \ref{tab7}. Here, we present the signal event rates using the pre-selection cuts $p_T(b_1/b_2)>20/20~\text{GeV},~\text{GeV},~p_T(e/ \mu)=10/8~\text{GeV}$. Notably, two mass observables, $m_T^H$ and $\Delta m_h$, are effective at significantly reducing background events. Additionally, while the signal event count is lower with the 20/20 GeV $b$-jet pre-selection, this cut nearly eliminates background events, ultimately improving the sensitivity to the signal.

	\begin{table}[!h]
	\begin{center}
		\resizebox{0.97\textwidth}{!}{
			\begin{tabular}{|c||c|c|c|c|c|c|c|c|c|c|} 
				\hline 
				BP  &   BP1 & BP2 & BP3 & BP4 & BP5 & BP6 & BP7 & BP8 & BP9 & BP10\\
				\hline \hline
				$m_h~\text{(GeV)}$& 17.67 & 25.9 & 28.56 & 33.20 & 37.56 & 40.68 & 47.27 & 54.03 & 43.44 & 49.39 \\
				\hline
				NoE($\mathcal{L},\sigma$)& 912.86 & 727.65& 687.432& 573.3&771.74& 769.18&1086.62& 1528.8  & 900.000 & 771.750 \\
				\hline \hline
				$e^\pm\mu^\mp$ & 156.934 & 151.874 &141.094 &114.84&146.44 & 136.2 &160.54 & 204.94 & 151.226 & 111.163 \\
				\hline 
				$m_Z$-veto &156.934 & 151.874 &141.094 &114.84&146.44 & 136.2 &160.54 & 204.94& 151.226 & 111.163 \\
				\hline
				2 $b$-jets & 13.6 & 20.38 & 19.02 & 15.3 & 17.86 & 16.02 & 18.34 & 23.7 & 17.39 & 11.06\\
				\hline
				$65~\text{GeV}< m_T^{H} < 125~\text{GeV}$ & 2.68 & 7.16 & 6.84 & 5.72 & 6.84 & 6.16 & 6.14 & 11.38 & 6.80& 4.19\\
				\hline
				$\Delta m_{h}<0.5$ & 1.86 & 5.5 & 5.56 & 4.5 & 5.32 & 4.8 & 5.54 & 8.2 & 5.25 & 3.13\\
				\hline
				$m_{T}^{ll}<62.5$ GeV & 1.86 & 5.5 & 5.56 & 4.48 & 5.32 & 4.8 & 5.52 & 8.2 & 5.25& 3.13\\
				\hline 
				$m_{bb}<62.5$ GeV & 1.86 & 5.5 & 5.56 & 4.48  & 5.32 & 4.78 & 5.52 & 8.12 & 5.23 & 3.13\\
				\hline 
		\end{tabular}}
		\caption{Event rates of the signal with $\sqrt{s}=13$ TeV and integrated luminosity 300 fb$^{-1}$ for different BPs are shown as a function of our cutflow. }
		\label{tab6}
	\end{center}
\end{table}
	\begin{table}[h!]
	\begin{center}
		\resizebox{0.87\textwidth}{!}{
			\begin{tabular}{|c||c|c|c||c|c|c||}
				\hline
				Process & \multicolumn{3}{c|}{$Zb\overline{b}$}& \multicolumn{3}{c|}{$t\overline{t}$}  \\
				\hline 
				NoE($\mathcal{L},\sigma$) &  \multicolumn{3}{c|}{2562000}  &  \multicolumn{3}{c|}{117600} \\
				\hline 
				$p_T(b_1/b_2)~\text{(GeV)}$ & 15/10 & 20/15 & 20/20& 15/10 & 20/15 & 20/20 \\
				\hline
				$e^\pm\mu^\mp$ & 15836.8 & 15836.8 &  15836.8 &  61413.5  &61413.5 & 61413.5 \\
				\hline
				$m_Z$-veto & 15801.4  & 15801.4&  15801.4& 54511.6 & 54511.6 &  54511.6\\
				\hline
				2 $b$-jets & 1512.57 & 1059.63 & 503.558 & 16871.4 & 13778.6 & 8843.26 \\
				\hline
				$65~\text{GeV}< m_T^{H} < 125~\text{GeV}$ & 272.439 & 154.314 & 33.2724 & 35.2954 &18.8916 & 3.087 \\
				\hline
				$\Delta m_{h}<0.5$ GeV & 117.072 & 30.0678  & -& 17.5266 & 7.6678& -\\
				\hline
				$m_{T}^{ll}<62.5$ GeV &  117.072 & 30.0678 & -& 14.2366 & 6.125 &-\\
				\hline
				$m_{bb}<62.5$ GeV & 117.072 & 30.0678 & -& 14.2366 & 6.125 &-\\
				\hline
		\end{tabular}}
		\captionof{table}{Event rates of the two dominant background processes with $\sqrt{s}=13$ TeV and integrated luminosity 300 fb$^{-1}$ as a function of our cutflow. }
		\label{tab7}
	\end{center}
\end{table}
Tab.~\ref{tgg} shows the significances after applying the kinematic cuts, for 300 fb$^{-1}$ and 3000 fb$^{-1}$ of each BP are shown. As previously mentioned, the 20/20 GeV pre-selection cut provides better significance and is recommended for the actual analysis. While it may not be possible to discover or exclude all BPs at LHC Run 3, they are fully within reach of the HL-LHC.

\begin{table}
	\begin{center}
		\resizebox{0.87\textwidth}{!}{
			\begin{tabular}{|c||c|c|c||c|c|c||}
				\hline
				\multirow{2}{*}{BP}& \multicolumn{3}{c|}{Significance ($\Sigma$), $\mathcal{L}=300~\text{fb}^{-1}$} & \multicolumn{3}{c|}{Significance ($\Sigma$), $\mathcal{L}=3000~\text{fb}^{-1}$} \\ 
				\cline{2-7}
				&    15/10 (GeV) & 20/15 (GeV) &  20/20 (GeV) & 15/10 (GeV) & 20/15 (GeV) &  20/20 (GeV)  \\
				\hline
				BP1  &   0.68   &  0.81 & 1.36    &  2.15    & 2.56   &  4.30    \\
				\hline
				BP2  &  1.30    &  1.64  & 2.34  & 4.11    &  5.18   &  7.39   \\
				\hline
				BP3  &  1.24     &  1.57 & 2.35  & 3.92  &   4.96  & 7.43  \\
				\hline
				BP4  &   1.07 &  1.32 & 2.11    &  3.38  &   4.17   & 6.67 \\
				\hline
				BP5  &  1.33    &  1.57 &   2.3   & 4.20 &  4.96   &  7.27 \\
				\hline
				BP6 &  1.22  &   1.44 & 2.18  &  3.85   & 4.55   & 6.89   \\
				\hline
				BP7  &  1.48    &  1.71  & 2.34     &  4.68 & 5.40   & 7.39  \\
				\hline
				BP8  & 2.14    &  2.37 & 2.84    &  6.76  & 7.49      & 8.9 \\
				\hline 
				BP9 & 1.36 &1.59 & 2.28 &4.3 & 5.02 & 7.2 \\
				\hline 
				BP10 & 1.0 & 1.11 & 1.76 & 3.16 & 3.51 & 5.56\\
				\hline 
		\end{tabular}}
		\captionof{table}{Significances ($\Sigma = \mathcal{N}_S/\sqrt{\mathcal{N}_S+\mathcal{N}_B}$) for our signal against the two dominant backgrounds
			with $\sqrt{s}=13$ TeV and integrated luminosity
			$300~\text{fb}^{-1}$ (left) as well as $3000~\text{fb}^{-1}$ (right).
			The pre-selection cuts are as given in}
		\label{tgg}
	\end{center}
\end{table}
\section{Conclusion}
 
Within the 2HDM Type-I, we have demonstrated that  both Run 3 of the LHC and the HL-LHC phase have the potential to probe the signal $gg \to H \to hh \to b\bar{b}\tau\tau$, especially in the presence of very low mass trigger thresholds (on the electrons and muons). By exploring several benchmark points (BPs) after scanning the 2HDM Type-I parameter space, we provide examples that can be further investigated by LHC collaborations. Additionally, a future increase in LHC collision energy from 13 TeV to 14 TeV could enhance the signal production rate by 10\%, further strengthening the reach of this analysis.

\section*{Acknowledgements}
 SM is supported in part through the NExT Institute and the STFC Consolidated Grant  ST/L000296/1.
CHS-T(SS) is supported in part(full) through the NExT Institute. YW’s work is supported by the Natural Science Foundation of China Grant No. 12275143, the Inner Mongolia Science Foundation Grant No. 2020BS01013 and the Fundamental Research Funds for the Inner Mongolia Normal University Grant No. 2022JBQN080. QSY is supported by the Natural Science Foundation of China under the Grants No. 11875260 and No. 12275143.

\bibliographystyle{unsrt}
\bibliography{bibio2}

\begin{thebibliography}{10}

\bibitem{ATLAS:2022vkf}
{A detailed map of Higgs boson interactions by the ATLAS experiment ten years
  after the discovery}.
\newblock {\em Nature}, 607(7917):52--59, 2022.
\newblock [Erratum: Nature 612, E24 (2022)].

\bibitem{CMS:2022dwd}
Armen Tumasyan et~al.
\newblock {A portrait of the Higgs boson by the CMS experiment ten years after
  the discovery}.
\newblock {\em Nature}, 607(7917):60--68, 2022.

\bibitem{CMS:2024uru}
Aram Hayrapetyan et~al.
\newblock {Search for exotic decays of the Higgs boson to a pair of
  pseudoscalars in the $\mu\mu$bb and $\tau\tau$bb final states}.
\newblock {\em Eur. Phys. J. C}, 84(5):493, 2024.

\bibitem{Eriksson:2009ws}
David Eriksson, Johan Rathsman, and Oscar Stal.
\newblock {2HDMC: Two-Higgs-Doublet Model Calculator Physics and Manual}.
\newblock {\em Comput. Phys. Commun.}, 181:189--205, 2010.

\bibitem{Bechtle:2020pkv}
Philip Bechtle, Daniel Dercks, Sven Heinemeyer, Tobias Klingl, Tim Stefaniak,
  Georg Weiglein, and Jonas Wittbrodt.
\newblock {HiggsBounds-5: Testing Higgs Sectors in the LHC 13 TeV Era}.
\newblock {\em Eur. Phys. J. C}, 80(12):1211, 2020.

\bibitem{Bechtle:2020uwn}
Philip Bechtle, Sven Heinemeyer, Tobias Klingl, Tim Stefaniak, Georg Weiglein,
  and Jonas Wittbrodt.
\newblock {HiggsSignals-2: Probing new physics with precision Higgs
  measurements in the LHC 13 TeV era}.
\newblock {\em Eur. Phys. J. C}, 81(2):145, 2021.

\bibitem{Mahmoudi:2008tp}
F.~Mahmoudi.
\newblock {SuperIso v2.3: A Program for calculating flavor physics observables
  in Supersymmetry}.
\newblock {\em Comput. Phys. Commun.}, 180:1579--1613, 2009.

\bibitem{Alwall:2014hca}
J.~Alwall, R.~Frederix, S.~Frixione, V.~Hirschi, F.~Maltoni, O.~Mattelaer,
  H.~S. Shao, T.~Stelzer, P.~Torrielli, and M.~Zaro.
\newblock {The automated computation of tree-level and next-to-leading order
  differential cross sections, and their matching to parton shower
  simulations}.
\newblock {\em JHEP}, 07:079, 2014.

\bibitem{Sjostrand:2006za}
Torbjorn Sjostrand, Stephen Mrenna, and Peter~Z. Skands.
\newblock {PYTHIA 6.4 Physics and Manual}.
\newblock {\em JHEP}, 05:026, 2006.

\bibitem{deFavereau:2013fsa}
J.~de~Favereau, C.~Delaere, P.~Demin, A.~Giammanco, V.~Lema\^\i{}tre,
  A.~Mertens, and M.~Selvaggi.
\newblock {DELPHES 3, A modular framework for fast simulation of a generic
  collider experiment}.
\newblock {\em JHEP}, 02:057, 2014.

\bibitem{Arhrib:2023apw}
A.~Arhrib, S.~Moretti, S.~Semlali, C.~H. Shepherd-Themistocleous, Y.~Wang, and
  Q.~S. Yan.
\newblock {Searching for
  H\textrightarrow{}hh\textrightarrow{}bb\textasciimacron{}\ensuremath{\tau}\ensuremath{\tau}
  in the 2HDM type-I at the LHC}.
\newblock {\em Phys. Rev. D}, 109(5):055020, 2024.

\end{thebibliography}

\end{document}